\documentclass[%
 reprint,
 amsmath,amssymb,
pra,
]{revtex4-2}

\usepackage{graphicx}% Include figure files
\usepackage{stfloats}
\usepackage{dcolumn}% Align table columns on decimal point
\usepackage{bm}% bold math
\usepackage{multirow}
\usepackage{graphics}
\usepackage{xcolor}
\usepackage{amssymb}
\usepackage{braket}
\usepackage{longtable}
\usepackage{rotating}
\usepackage{diagbox}

\usepackage{nicefrac}

\usepackage{scalerel}
\usepackage[normalem]{ulem}

\newcommand{\ME}[3]{ \mbox{$\langle #1\,|\,#2\,|\,#3\rangle $} }
\newcommand{\MEred}[3]{ \mbox{$\langle #1\,||\,#2\,||\,#3\rangle $} }
\newcommand{\CGC}[6]{ \mbox{$ ( #1 #2 , #3 #4\,|\, #5 #6 ) $} }
\newcommand{\SechsJ}[6]{ \mbox{$ %
            \arraycolsep0.25ex %
            \left\{ \begin{array}{ccc} %
                       #1 & #2 & #3 \vspace{0.5ex}\\%
                       #4 & #5 & #6 %
                   \end{array} \right\} $} }

\begin{document}

\preprint{APS/123-QED}

%1. "Analyzing the Benefits of Polarization Control in the RABBITT Setup"
%2. "An Analysis of the Advantages of Polarization Control in the RABBITT Setup"
%3. "The Advantages of Polarization Control in RABBITT: An Analytical Study" 

\title{The Advantages of Polarization Control in RABBITT}

\author{Maria M. Popova}
 \affiliation{ 
Skobeltsyn Institute of Nuclear Physics, Lomonosov Moscow State University, 119991 Moscow, Russia
}
 \affiliation{ A.V. Gaponov-Grekhov Institute of Applied Physics, Russian Academy of Sciences, 603950 Nizhny Novgorod, Russia}

\author{Sergei N. Yudin}
 \affiliation{ 
Skobeltsyn Institute of Nuclear Physics, Lomonosov Moscow State University, 119991 Moscow, Russia
}

\author{Alexei N. Grum-Grzhimailo}
\affiliation{ 
Skobeltsyn Institute of Nuclear Physics, Lomonosov Moscow State University, 119991 Moscow, Russia
}

\author{Elena V. Gryzlova}
\affiliation{ 
Skobeltsyn Institute of Nuclear Physics, Lomonosov Moscow State University, 119991 Moscow, Russia
}
 \affiliation{ A.V. Gaponov-Grekhov Institute of Applied Physics, Russian Academy of Sciences, 603950 Nizhny Novgorod, Russia}
\email{gryzlova@gmail.com}

\date{\today}

\begin{abstract}
The RABBITT setup is theoretically studied for various combinations of  XUV and IR field components polarization: 'linear+linear', `linear+circular' with crossed propagation directions, and `circular+circular' with parallel propagation directions. The general properties of photo\-electron angular distributions and  their responses to the variation of the IR pulse delay are studied. Numerical simulations  are performed for the  neon valence shell ionization into the region of structure\-less continuum using two approaches based on time dependent perturbation theory and solution of  rate equations. To distinguish between "geometrical" governed by fields' polarization and spectroscopic features, additional analysis for the case of $s$-shell ionization is presented.
\end{abstract}


\maketitle

%%%%%%%%%%%%%%%%%%%%%%%%%%%%%%%%%%%%%%%%%%

Since the very beginning of photo\-ionization experiments, it has been well known that angle-resolved measurements provide more profound and detailed information about a process than measurements of the angle-integrated probabilities \cite{Becker}. Application of a multicolor field significantly enhances experimental capabilities because the polarization and propagation directions of the field components can be modulated separately \cite{Wuilleumier2006}. That paved a way to the measurements of different types of dichroism, primarily linear and circular magnetic dichroism. The first attempts to access dynamical peculiarities of multiphoton ionization were based on variation of the time offset (lag)  between pump and probe fields \cite{PhysRevA.80.025402,10.1063/1.2716360}. The development of highly coherent sources of extreme ultraviolet (XUV)  and X-ray radiation, such as high-order harmonic generation (HHG) setups \cite{Lewenstein1994,Sansone2006,strelkov2016} or  X-ray free electron lasers (XFELs) \cite{Prince2016,CALLEGARI20211}, opens a fruitful perspective for the vector correlation control in the atto\-second metrology. That has made possible experimental studying of the dynamics of small quantum systems on atto\-second timescales 
\cite{Krausz2009,Pazourek2013,Vos2018,Ossiander2018}. 

Atto\-second metrology based on the RABBITT (Reconstruction of Atto\-second Beating By Interference of Two-photon Transitions) scheme \cite{Veniard1996}, when an electron is promoted to the continuum by an XUV harmonic and then additionally absorbs or emits an optical (IR) photon, started with angle-integrated experiments \cite{Paul2001,Muller2002,Mairesse2003,Dahlstrom2012} and appropriate theoretical considerations \cite{DAHLSTROM201353,Harth2019,Kheifets2021}. Meanwhile, the path\-ways' interference that underlies the scheme manifests differently in different waves ($s\,,p\,,d$...) at the different photo\-emission angles \cite{Laurent2012}. The RABBITT experiments have advanced to the angle-resolved  case \cite{Aseyev2003,villeneuve2017,Joseph2020,Cirelli2018,Jiang}, and different theoretical approaches have been applied \cite{Kheifets2023}. Angle-resolved measurements provide more detailed information and allow for the separation different pathways \cite{Heuser2016,PhysRevLett.109.083001,Hockett2017,Busto2019,Fuchs2020,Bharti2023,Peschel2022}. Moreover, keeping in mind that energy dependency of vector correlations differs from  that of  angle-integrated one \cite{Grum2005} and may be narrower and shifted, the investigation may be useful for resolving  overlapping resonances or in case when the XUV spectrum is very broad, causing single- and two-photon signals overlapping in energy.

The possibility to perform angle-resolved measurements is a milestone for schemes with a mixture of waves of different parities in a side\-band: (a) XUV harmonics differ by $3\omega$ \cite{Maroju2020,Maroju_2021,Maroju2021}, (b) a bi\-chromatic combination $\omega$ and $2\omega$ is applied \cite{villeneuve2017}, and (c)  significant quadrupole effects are expected.

In spite of great progress occurred in the field, very few investigations involving different harmonics' polarizations (and their directions)  are reported \cite{Hockett2017,Boll2017,Kheifets2023,Liao2024}. As progress in the generation of circularly and elliptically polarized harmonics is essential \cite{Fleischer,Ivanov2014,Han,Sarantseva2023}, there is a need for a systematic investigation of polarization effects in the RABBITT setup. 
Special attention in such an investigation must be paid to angle-resolved observables. It is important to emphasize that for some combinations of the components' polarizations and propagation directions, RABBITT oscillations may appear only in angle-resolved parameters, even in a conventional scheme with odd XUV components that differ by $2\omega$. 

Unless otherwise specified, the atomic system of units is used. 

\section{Theoretical basement for the RABBIT description}

In the paper we extend the approaches based on the solving of rate equations and time dependent perturbation theory applied earlier for linearly polarized fields \cite{popova2023,yudin2023} to the systems with more complex geometries (polarization and propagation direction of the field component). Thus, here we briefly describe the methods clearly indicating polarization aspects.

The electromagnetic field is presented as a sum of XUV harmonics of an order $N$ generated on a seed IR pulse:
\begin{align}\label{eq:field}
\bm{E}(t)&=\Re\Big{[}\sum_{N\Lambda\lambda}E_{\rm{xuv}}c_\Lambda\bm{\epsilon}_{\Lambda} e^{-i(N\omega t+\phi_{N})}+\nonumber\\
&\hspace{90pt}E_{\rm{ir}}c_\lambda\bm{\epsilon}_{\lambda}e^{-i(\omega t+\phi)}\Big{]},
\end{align}
where $E_{\rm{xuv}}=E_{\rm{xuv}}^0\cos^2(\frac{t}{\tau})$ and $E_{\rm{ir}}=E_{\rm{ir}}^0\cos^2(\frac{2t}{\tau})$ %E_{\rm{xuv},N}(t)=\cos^2(\frac{t}{\sqrt{2}\sigma_{\rm{xuv}}})$ 
are slowly varying envelopes with $E_{\rm{ir}}^0$ and $E_{\rm{xuv}}^0$ being strengths of the IR and XUV components, respectively, and $\tau$ determines the pulse duration; $\phi_{N}$ is Nth XUV component phase, and $\phi$ is the varying phase shift of the IR pulse connected with the conventional in such experiments IR pulse delay $\tau_{\rm del}$ as  $\phi=2\omega \tau_{\rm del}$.
We use cosine envelope instead of Gaussian one because it provides smoothness at the (finite) edges of the pulse. Polarization of the field is determined by decomposition  over cyclic coordinate vectors $\bm{\epsilon}_{\lambda/\Lambda=0,\pm1}$ with coefficients $c_{\lambda/\Lambda}$. The coefficients before $\bm{\epsilon}_{\lambda/\Lambda=0}$ appear when photon wave vector  is not parallel to the quantization axis $z$ (see Table 1 below).

The atomic Hamiltonian is presented in the form:
\begin{equation} \label{eq:sheq}
i \, \frac{\partial}{\partial t} \Psi({\bm r},t) = \left(\hat{H}_{\rm at} + \hat{H}_{\rm int}(t) \right) 
\Psi({\bm r},t)\,,
\end{equation}
where $\hat{H}_{\rm at}$ is unperturbed Hamiltonian and $\hat{H}_{\rm int}(t)$ describes the interaction with electromagnetic field in the dipole approximation and  velocity gauge with electric field potential $\bm{A}(t)=-c\int \bm{E}(t)dt$.

In the $LS$-coupling scheme within the frozen core approximation, eigen\-functions of the system $\psi_{\alpha_n}(\varepsilon_n,{\bm r})$ depend on the following quantum numbers: energy $\varepsilon_n$ ($\varepsilon$ without an index if a state belongs to a continuum spectrum), core (ion) orbital momentum and spin $L_c$ and $S_c$, active electron angular momentum $l$ and spin $s=\nicefrac{1}{2}$, total angular momentum $L$ and spin $S$, and their projections $M_L=M$ and $M_S$.  
 Accounting that the electric dipole operator does not change spin $\Delta S=0$ and  an atom is initially in  a state with a definite spin, we can rule out the spin quantum numbers for brevity: $\psi_{\alpha_n}(\varepsilon_n,{\bm r})\equiv\psi_{(L_cl)L(S_c\frac{1}{2})SM_LM_S}(\varepsilon_n,{\bm r})\equiv\psi_{(L_cl)LM}(\varepsilon_n,{\bm r})$. A wave function of the system $\Psi({\bm r}, t)$ is expanded in the basis of eigen\-functions of the unperturbed Hamiltonian:
\begin{align} \label{eq:wf_decomp}
&\hat{H}_{\rm at} \psi_{\alpha_n}{(\varepsilon_{n},{\bm r})} = \varepsilon_{n}\psi_{\alpha_n}{(\varepsilon_{n},{\bm r})},\\
&\Psi({\bm r}, t) = \sum_{L_clLM} \hspace{-5pt}\Big{(} \sum_{n} \mathcal  U_{(L_cl)LM}(\varepsilon_n,t) \psi_{\alpha_n}{(\varepsilon_{n},{\bm r})} e^{-i  \varepsilon_n t}\nonumber\\
&\hspace{50pt}+\int \hspace{-3pt}d\varepsilon \,   \mathcal  U_{(L_cl)LM}(\varepsilon,t)\psi_{\alpha_\varepsilon}{(\varepsilon,{\bm r})}
e^{-i  \varepsilon t} \Big{)}\,,
\end{align}
where $\alpha_\varepsilon$ means a set of quantum numbers of a state with an electron in continuum.

Then the system of differential equations for expansion coefficients:
\begin{eqnarray}
     \label{eq:diff}
 &&\hspace{-9pt}\frac{d   \mathcal  U_{(L_cl)LM}(\varepsilon_{n'},t)}{dt} =\nonumber\\
 &&\hspace{-9pt}-i \hskip1truemm \int\hskip-5truemm \sum_{n} e^{i(\varepsilon_{n'}-\varepsilon_{n})t} \ME{\psi_{\alpha_n'}}{\hat{H}_{\rm int}(t)}{\psi_{\alpha_n}} \,   \mathcal  U_{(L_cl)LM}(\varepsilon_{n},t),
 \end{eqnarray}
is solved numerically in the {\sl Rate Equations (RE)} method (for the continuum part $\varepsilon_n\rightarrow\varepsilon$ and $\psi_{\alpha_n} \rightarrow \psi_{\alpha_\varepsilon}$).
To describe the states of the continuum in (\ref{eq:diff}), the  discretization of continuum was applied, i.e. the integration was replaced by summation with even energy step $d\varepsilon=10^{-3}$~a.u. %, which is good for describing relatively short energy spans ($\approx 10$ eV) with equal quality.
Thereby $|  \mathcal  U_{(L_cl)LM}(\varepsilon_{\epsilon},t) |^2$ is the probability to find an electron in the neighborhood $d\epsilon$ of the energy value $\varepsilon$ at time $t$.

The decomposition of the vector potential into cyclic coordinates can be further decomposed into ${A}_{\rm xuv}(t)$ that describes XUV comb and ${A}^u_{\rm ir}(t)$ and ${A}^d_{\rm ir}(t)$ that describe IR and behave as $e^{-i(\omega t+\phi)}$ and $e^{i(\omega t+\phi)}$, respectively. The component ${A}^u_{\rm ir}(t)$ is responsible for an absorption of an IR photon, and ${A}^d_{\rm ir}(t)$ --- its emission.
 
Within the framework of {\sl perturbation theory (PT)}, the coefficients $\mathcal U_{(L_cl)LM}(\varepsilon_{n},t)$ in the equation ($\ref{eq:diff}$) are in turn expanded into the series. 

Henceforth, we specify a target as a shell of an unpolarized atom with initial orbital momentum $L=0$. 

The first order coefficients describe direct ionization  to the main photo\-lines (ML) by XUV components of the electric field:
\begin{align}\label{eq:ampl1}
& \mathcal  U^{(1)}_{(L_cl)LM}(\varepsilon_f,t)=c_\Lambda\frac{1}{\sqrt{3}}\CGC{0}{0}{1}{\Lambda}{1}{M}D^{(1)}_{(L_cl)1}\,,\\
&D^{(1)}_{(L_cl)1}=-i\MEred{\varepsilon_{f};(L_cl)1}{\hat{D}}{\varepsilon_{0},\!0}\int_{-\tau/2}^{\tau/2} \hspace{-10pt} A_{\rm xuv}(t)e^{i(\varepsilon_f-\varepsilon_{0})t}dt\,,
\end{align}
where  $\CGC{0}{0}{1}{\Lambda}{1}{M}$ is equal to 1 for each $\Lambda$ but it determines a specific $M$, and we suppose the component to be completely polarized that means that with appropriate choice of a coordinate system only one $c_\Lambda\neq0$. 

The second order amplitudes describe absorption (`$u$') or emission (`$d$') of an IR photon leading to appearance of side\-bands (SB) by up- and down-energy transitions:
\begin{align}\label{eq:ampl2}
& \mathcal  U^{(2),u/d}_{(L_cl)LM}(\varepsilon_f,t)=\frac{(\pm1)^\lambda}{\sqrt{3}\hat{L}}\sum_{\lambda}c_{\lambda}\CGC{1}{\Lambda}{1}{\pm\lambda}{L}{M}D^{(2),u/d}_{(L_cl)L}\,,\\
\label{eq:amplD2}
&D^{(2),u/d}_{(L_cl)L}=  
\hspace{5pt}\int \hskip-5.7truemm \sum_{n}\!\!\MEred{\varepsilon_f,\!(L_cl)L}{D}{\varepsilon_{n},\!1}\MEred{\varepsilon_{n},\!1}{D}{\varepsilon_{0},\!0}\nonumber\\
&\hspace{7pt}\int_{-\tau}^{\tau}A_{\rm IR}^{u/d}(t)e^{i(\varepsilon_f-\varepsilon_{n})t} \int_{-\tau/2}^{t}\hspace{-10pt} A_{\rm xuv}(t')e^{i(\varepsilon_{n}-\varepsilon_{0})t'}dt'dt\,,
\end{align}
where '$+$' sign is for absorption amplitude, and '$-$' sign is for emission. Factor $(\pm1)^\lambda$ comes from taking complex conjugation in cyclic basis: $\epsilon_{+1}^*=-\epsilon_{-1}$ and vice versa.
 %Here $c_{\lambda_{\rm ir}=0}=1$ for linear polarization in $z$ direction, $c_{\lambda_{\rm ir}=\pm1}=-\lambda/\sqrt{2}$ for linear polarization in $x$ direction, $c_{\lambda_{\rm ir}=1}=\pm1$ for absorption/emission of right circular photon, and $c_{\lambda_{\rm ir}=-1}=\pm1$ for absorption/emission of left circular photon.
 In equation~(\ref{eq:ampl2}), conventional notation for Clebsch---Gordan coefficients is used and $\hat{a}=\sqrt{2a+1}$.
 % Note that the $`u/d'$ dependence of (\ref{eq:amplD2}) lies in final photoelectron energy $\varepsilon_f$.

The photo\-electron angular distribution (PAD) in PT and RE is described as:
\begin{align}\label{eq:W}
&W(\varepsilon_f,t;\vartheta,\varphi)=\frac{1}{4\pi}\sum_{kq ll'LL'\atop nn'\nu\nu'}(-1)^{L_c+L+L'+k-M'}\hat{l}\hat{l}'\hat{L}\hat{L}'\nonumber\\
&\hspace{35pt}\CGC{l}{0}{l'}{0}{k}{0}\CGC{
L}{M}{L'}{-M'}{k}{q}\SechsJ{l}{L}{L_c}{L'}{l'}{k}\nonumber\\
&\hspace{10pt} \mathcal  U^{(n),\nu}_{(L_cl)LM}(\varepsilon_f,t) \mathcal  U^{(n'),\nu'\ast}_{(L_cl')L'M'}(\varepsilon_f,t) \frac{\sqrt{4\pi}}{\hat{k}}Y_{kq}(\theta,\varphi) \,,
\end{align}
where $\nu$ is an order of amplitude in PT (in RE the resulting amplitude $\mathcal  U_{(L_cl)LM}(\varepsilon_f,t)$ is an `infinite sum' over $\nu$). In equation (\ref{eq:W}), conventional notations for Wigner $6j$-symbol and spherical harmonics are used.

In eq. (\ref{eq:W}), there are terms corresponding to the absorption (`$\mathcal U^{(2),u}\mathcal U^{(2),u\ast}$') and emission (`$\mathcal U^{(2),d}\mathcal U^{(2),d\ast}$') of IR photon which do not depend on the IR time-delay; and interference between absorption and emission amplitudes (`$\mathcal U^{(2),u}\mathcal U^{(2),d\ast}$') which depends on the IR delay and oscillates on the double IR frequency $2\omega$.

\section{\label{sec:1} The PAD for different geometries}

\begin{figure}
\begin{minipage}[c]{\textwidth}
\includegraphics[width=0.95\textwidth]{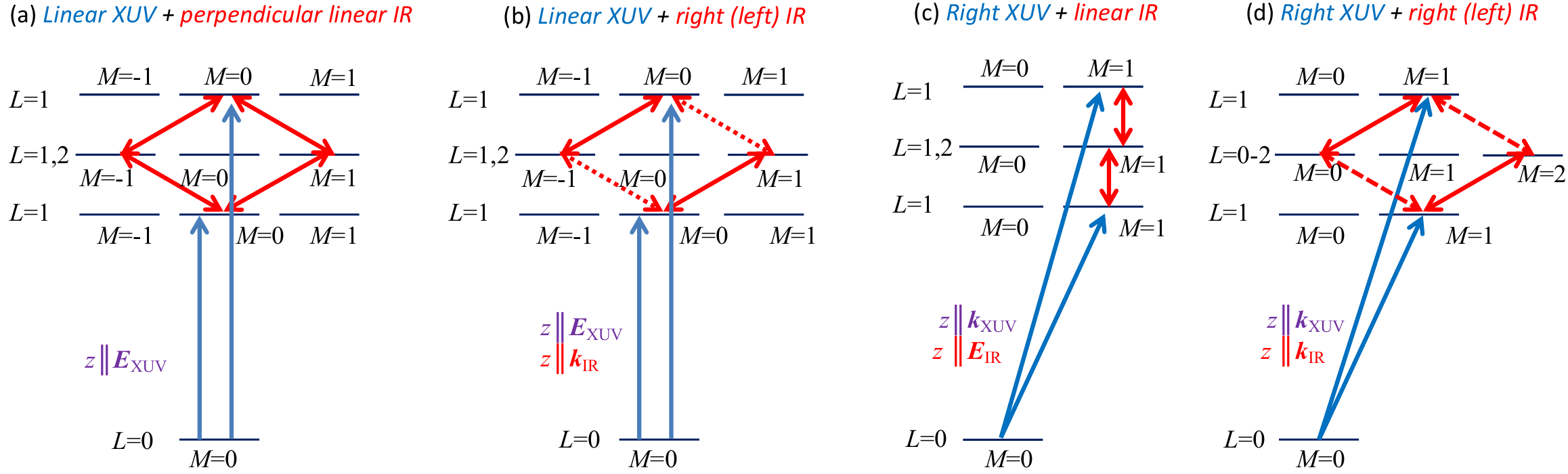} 
\caption{The scheme of RABBITT for different geometries: (a) XUV and IR are linearly polarized in the orthogonal directions; (b) Circularly polarized IR and linearly polarized XUV harmonics; (c) Circularly polarized XUV and linearly polarized IR harmonics; (d) both IR and XUV harmonics are circularly polarized.}
\label{fig:1}
\end{minipage}
\end{figure}

In figure (\ref{fig:1}), there are schemes of RABBITT spectroscopy for the systems under consideration: (a)  XUV and IR are linearly polarized in the orthogonal directions (quantization axis is along the XUV  component  polarization  $z\parallel \bm{E}_{\rm{xuv}}$); (b) Circularly polarized IR and linearly polarized ($\bm{E}_{\rm{xuv}}\parallel{\bm k }_{\rm{IR}}$)  XUV  (quantization axis is along the XUV  component  polarization $z\parallel \bm{E}_{\rm{xuv}}$); (c) Circularly polarized XUV and linearly polarized IR (${\bm E}_{\rm{IR}}\parallel {\bm k}_{\rm{xuv}}$) harmonics (quantization axis is along the IR component polarization $z\parallel \bm{E}_{\rm{IR}}$); (d) both IR and XUV harmonics are circularly polarized (quantization axis is along the field propagation direction $z\parallel \bm{k}_{\rm{IR}}\parallel \bm{k}_{\rm{XUV}}$). 

% \begin{figure*}
% \centering
% \includegraphics[width=0.95\textwidth]{Figures/Figure 1_v4.pdf} 
% \caption{The scheme of RABBITT for different geometries: (a) XUV and IR are linearly polarized in the orthogonal directions; (b) Circularly polarized IR and linearly polarized ($\bm{E}|| \bm{k}_{\rm{IR}}$) XUV harmonics; (c) Circularly polarized XUV and linearly polarized IR ($\bm{E}_{\rm{IR}}|| \bm{k}$) harmonics; (d) both IR and XUV harmonics are circularly polarized.}
% \label{fig:1}
% \end{figure*}

\begin{table}[h!]
\centering
\caption{The polarization coefficients $c_{\lambda}$ (\ref{eq:field}) in the cyclic basis $\{-,0,+\}$ for different geometries and the list of allowed channels.}
\label{tab1}
\begin{tabular}{c|c|c|c|c}
%\noalign{\hrule height 1pt} %please confirm the change
geometries  &  (a) & (b) & (c) & (d) \\
%\noalign{\hrule height .5pt}\\
system & $z||\bm{E}_{\rm{xuv}}$ & $z||\bm{E}_{\rm{xuv}}$ & $z||\bm{k}_{\rm{xuv}}$ & $z||\bm{k}_{\rm{xuv}}$\\
\hline
$c_{\rm{xuv}}$ & $\{0,1,0\}$ & $\{0,1,0\}$ & $\{0,0,1\}$ & $\{0,0,1\}$\\
$c_{\rm{IR}}$ & $\{\frac{e^{i\gamma}}{\sqrt{2}}   ,0,-\frac{e^{-i\gamma}}{\sqrt{2}}\}$ & $\{0,0,1\}$ & $\{0,1,0\}$ &  $\{0/1,0,1/0\}$\\
\hline
Helium-like & $\varepsilon d^{u,d}$  & $\varepsilon d^{u,d}$ & $\varepsilon d^{u,d}$ & $\varepsilon d^{u,d}$, $\varepsilon s^{d}$ \\
\hline
 & - & - & - & $\varepsilon p^1S^{u}$ \\
 Noble gases & $\varepsilon p^1P^{u,d}$  & $\varepsilon p^1P^{u,d}$ & $\varepsilon p^1P^{u,d}$ & $\varepsilon p^1P^{u}$ \\
 & $\varepsilon p^1D^{u,d}$  &  $\varepsilon p^1D^{u,d}$ & $\varepsilon p^1D^{u,d}$ & $\varepsilon p^1D^{u,d}$ \\
 & $\varepsilon f^1D^{u,d}$  &   $\varepsilon f^1D^{u,d}$ & $\varepsilon f^1D^{u,d}$ &  $\varepsilon f^1D^{u,d}$ \\
  \multicolumn{5}{l}{ $^\ast$: the index $(u)$ marks the channels proceeding with } \\\multicolumn{5}{l}{absorption of the IR photon, $(d)$ with emission.}
\end{tabular}
\end{table}

The case of the linearly polarized in the same direction XUV and IR fields we studied earlier in \cite{yudin2023}.  The quantum numbers in figure (\ref{fig:1}) are depicted for a system with initial orbital momentum $L=0$ (a noble gas). Further we apply the approach to valence shell ionization of Neon and specify the general equations for the $s$-shell ionization ($1s$ of He, $2s$ of Ne, $3s$ of Ar etc). 

In table \ref{tab1}, channels allowed for the different geometries under consideration are presented. 
For the (a)--(c) geometries,  final projection of magnetic quantum number  $M=1$, therefore, the terms are $P\,(L=1)$ and $D\,(L=2)$. For the helium-like systems, there is a single $d$-wave channel because parity conservation rule prohibits emission of $\varepsilon p$-wave. The (d) geometry is exceptional because the $S\,,(L=0)$ (for neon) and $\varepsilon s$ (for helium) channel is involved for down-energy (equal helicities of IR and XUV) or up-energy (opposite helicities) transitions. 

\subsection{A shell of a noble gas}

It is convenient to extract a fields' polarization-independent dynamical parameter $B^{(\nu\mu)}_k[L,L']$ from the general angular distribution equation (\ref{eq:W}). For the side\-bands in the second order of PT it takes a form:
\begin{eqnarray}
B^{(\mu\mu')}_k[L,L']&=&\frac{(-1)^{L_f+L+L'}}{12\pi}\sum_{ll'}\hat{l}\hat{l}'\CGC{l}{0}{l'}{0}{k}{0}\, \label{eq:Buu}\\
&&\hspace{-10pt}\SechsJ{L}{L'}{k}{l'}{l}{L_f}D^{(2),\mu}_{(L_cl)L}D^{(2),\mu'\ast}_{(L_cl)L}\,,\nonumber
%D^{(2,\mu)}[(L_fl)L]D^{(2,\nu)}[(Lfl')L']^{\ast}\,,\nonumber
\end{eqnarray}
where $\mu\,,\mu'=u,d$, and the dynamical parameters obey permutation equation  $B^{(\mu\mu')}_k[L,L']=B^{(\mu'\mu)}_k[L',L]^{\ast}$. In the case of the RE, the definition (\ref{eq:Buu}) is fair, except that instead of the second-order term $D^{(2),\mu}_{(L_cl)L}$ one must consider a complete amplitude that possesses given quantum numbers. 

\begin{figure}
\begin{minipage}[c]{\textwidth}
\centering
\includegraphics[width=0.8\textwidth]{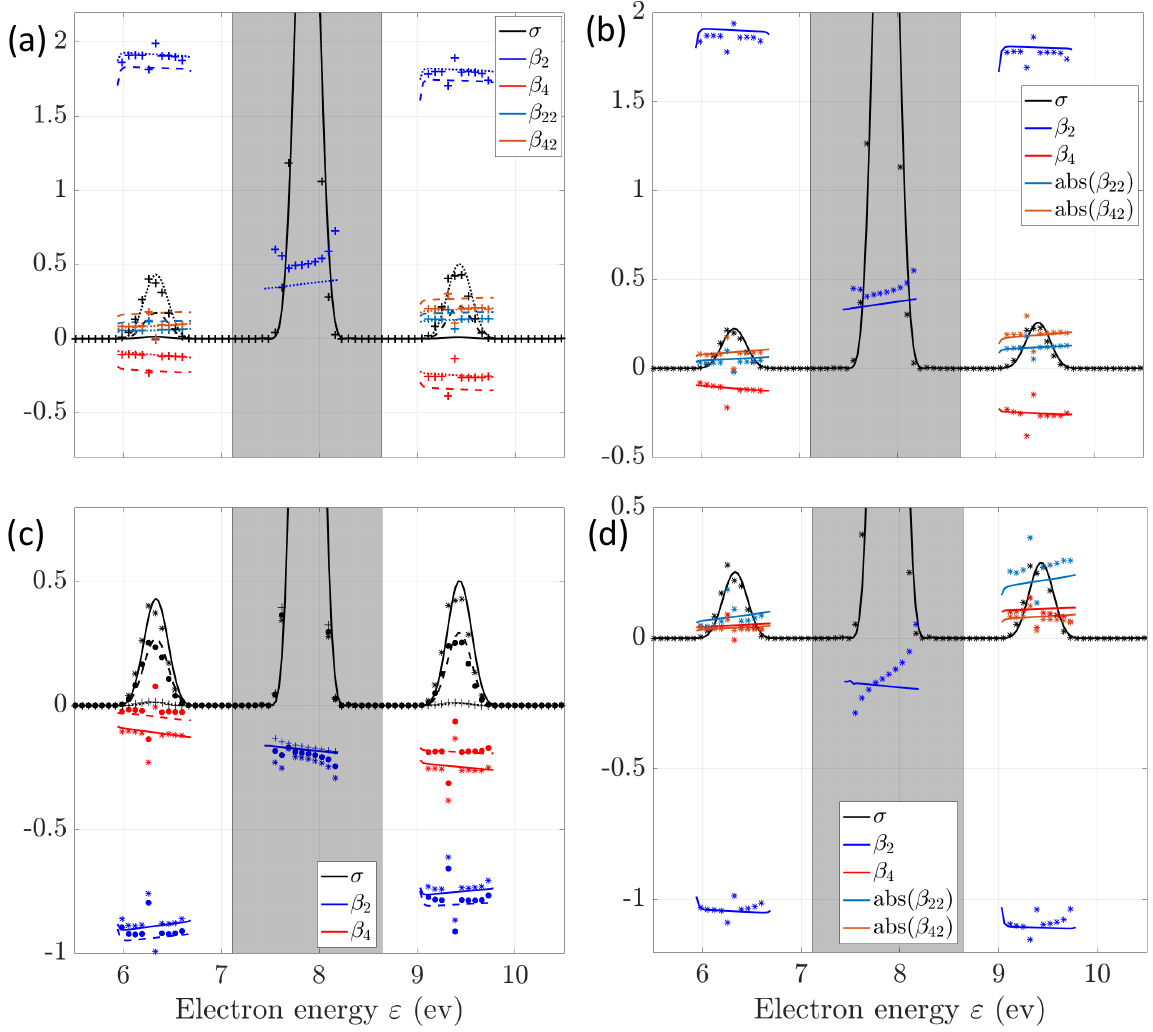} %в дискуссии дается ссылка на убранные ранее кривые
\caption{Spectra (in arbitrary units) and angular anisotropy parameters as defined in eqs. (\ref{eq:a})--(\ref{eq:d}) for geometries (a)--(d), respectively. Colors indicate a specific parameter (see the legends), and line type --- IR phase and a model (solid lines are for PT and $\phi=0$, stars are for RE and $\phi=0$; dashed lines and circles are for PT and RE for $\phi=\pi/4$, respectively; dotted lines and crosses are for PT and RE for $\phi=\pi/2$.}
\label{fig:results_neon}
\end{minipage}
\end{figure}

Now the angular distributions for specific geometries can be written down in an easier for an analysis  form:
\begin{widetext}
\begin{eqnarray}\label{eq:a}
W(\theta,\varphi)^{(a)}&=&\sum_{kLL'}\frac{(-1)^{L+L'+1}}{2}(B^{(dd)}_k[L,L']+B^{(uu)}_k[L,L']+B^{(ud)}_k[L,L']+B^{(du)}_k[L,L'])) \,\nonumber\\
 &&(\CGC{L}{1}{L'}{-1}{k}{0}P_{k}(\cos\theta)+(-1)^{L'}\CGC{L}{1}{L'}{1}{k}{2}\frac{\sqrt{4\pi}}{\hat{k}}(\eta Y_{k2}(\theta,\varphi)+\eta^{\ast} Y_{k-2}(\theta,\varphi))\\\label{eq:a2}
&=&\frac{\sigma^{(a)}}{4\pi}\left(1+\sum_{k=2.4} \beta^{(a)}_kP_{k}(\cos\theta)+\beta^{(a)}_{k2}\frac{\sqrt{4\pi}}{\hat{k}}\left(\eta Y_{k2}(\theta,\varphi)+\eta^{\ast} Y_{k-2}(\theta,\varphi)\right)\right)\,;\\
 %W(\theta,\varphi)^{(a)}&=&-\sum_{kLL'}
% \CGC{1}{0}{1}{1}{L}{1}\CGC{1}{0}{1}{1}{L'}{1}
 %(B^{(ud)}_k[L,L']+B^{(du)}_k[L,L'])\,\nonumber\\
 %&& \frac{\sqrt{4\pi}}{\hat{k}}(\CGC{L}{1}{L'}{-1}{k}{0}Y_{k0}(\theta,\varphi)+(-1)^{L'}\CGC{L}{1}{L'}{1}{k}{2}(\kappa Y_{k2}(\theta,\varphi)+\kappa^{\ast}Y_{k-2}(\theta,\varphi)  )\,.\\
 \label{eq:b}
 W(\theta,\varphi)^{(b)}&=&\sum_{kLL'}\frac{(-1)^{L+L'+1}}{2}\CGC{L}{1}{L'}{-1}{k}{0}(B^{(dd)}_k[L,L']+B^{(uu)}_k[L,L'])P_{k}(\cos\theta)\\
 &+&\CGC{L}{1}{L'}{1}{k}{2} \frac{(-1)^{L}}{2} 
 \frac{\sqrt{4\pi}}{\hat{k}}(B^{(ud)}_k[L,L']Y_{k2}(\theta,\varphi)+B^{(du)}_k[L',L]Y_{k-2}(\theta,\varphi))\,\nonumber\\
 \label{eq:b2}
 &=&\frac{\sigma^{(b)}}{4\pi}\left(1+\sum_{k=2.4} \beta^{(b)}_kP_{k}(\cos\theta)+\frac{\sqrt{4\pi}}{\hat{k}}\left(\beta^{(b)}_{k2} Y_{k2}(\theta,\varphi)+\beta^{(b)\ast}_{k2} Y_{k-2}(\theta,\varphi)\right)\right)\,;\\ 
 \label{eq:c}
 W(\theta,\varphi)^{(c)}&=& \sum_{kLL'}\CGC{L}{1}{L'}{-1}{k}{0} \frac{-1}{2}
(B^{(dd)}_k[L,L']+B^{(uu)}_k[L,L']+B^{(ud)}_k[L,L']+B^{(du)}_k[L',L])P_{k}(\cos\theta)\,;\\\label{eq:c2}
&=&\frac{\sigma^{(c)}}{4\pi}\left(1+\sum_{k=2,4} \beta^{(c)}_kP_{k}(\cos\theta)\right)\\
%W(\theta,\varphi)^{(c)}&=&-\sum_{kLL'}\CGC{L}{1}{L'}{-1}{k}{0} \CGC{1}{1}{1}{0}{L}{1}\CGC{1}{1}{1}{0}{L'}{1}(B^{(ud)}_k[L,L']+B^{(du)}_k[L',L])P_{k}(\cos\theta)\,.\\
 W(\theta,\varphi)^{(d)}&=&\sum_{kLL'}(\CGC{L}{0}{L'}{0}{k}{0}  \CGC{1}{1}{1}{-1}{L}{0}\CGC{1}{1}{1}{-1}{L'}{0}B^{(dd)}_k[L,L']+\CGC{2}{2}{2}{-2}{k}{0}B^{(uu)}_k[2,2])P_{k}(\cos\theta)\,\nonumber\label{eq:d}\\ 
&&-\CGC{2}{2}{L'}{0}{k}{2}\CGC{1}{1}{1}{-1}{L'}{0}\frac{\sqrt{4\pi}}{\hat{k}}(B^{(ud)}_k[2,L']Y_{k2}(\theta,\varphi)+B^{(du)}_k[L',2]Y_{k-2}(\theta,\varphi)) \,\\
\label{eq:d2}
 &=&\frac{\sigma^{(d)}}{4\pi}\left(1+\sum_{k=2.4} \beta^{(d)}_kP_{k}(\cos\theta)+\frac{\sqrt{4\pi}}{\hat{k}}\left(\beta^{(d)}_{k2} Y_{k2}(\theta,\varphi)+\beta^{(d)\ast}_{k2} Y_{k-2}(\theta,\varphi)\right)\right)\,.
%\textcolor{red}{ %W(\theta,\varphi)^{(d)}}&=&\textcolor{red}{-\sum_{kL'}\frac{\sqrt{4\pi}}{\hat{k}}(\CGC{2}{2}{L'}{0}{k}{2}\CGC{1}{1}{1}{-1}{L'}{0}B^{(ud)}_k[2,L']Y_{k2}(\theta,\varphi)}+\nonumber\\
%&&\textcolor{red}{\CGC{L}{0}{2}{-2}{k}{-2}\CGC{1}{1}{1}{-1}{L}{0}B^{(du)}_k[L',L]Y_{k-2}(\theta,\varphi)) } \,.
\end{eqnarray}
\end{widetext}
The parameter $\eta=-\exp[-2i\gamma]/2$ is defined by the angle $\gamma$ of IR polarization vector with respect to the $x$-axis in the $xy$-plane. Equations (\ref{eq:a2}), (\ref{eq:b2}), (\ref{eq:c2}) and (\ref{eq:d2}) themselves are the definition of the integral photo\-emission probability $\sigma^{a,b,c,d}$ and angular anisotropy parameters $\beta^{a,b,c,d}_{kq}$.  The parameters $\beta_{22(42)}$  are defined to be in consistency with conventional parameters $\beta_{2(4)}$ ($\frac{\sqrt{4\pi}}{\hat{k}}Y_{k0}(\theta,\varphi)$ is a Legendre polynomial $P_k(\cos\theta)$). The angular anisotropy parameters $\beta_{kq}$ allow us characterize the angular distributions by a few numbers (see Fig. \ref{fig:results_neon}). 
% The angular distributions are presented both in terms of dynamical parameters  $B^{(\nu\mu)}_k[L,L']$ which allows us to investigate symmetries and their dependence on IR-pulse delay, and 

%The systems (b)-(d) would possess of an axial symmetry for incoherent field's components, the system (a) would possess of three orthogonal symmetry planes. While for the systems (a) and (c) RABBITT oscillations may appear in the integral probabilities, for (b) and (d) the effect may be observed in the PAD only. Remarkable that the system (c) keeps an axial symmetry even for coherent harmonics. That's follows from the fact that only one projection ($M=1$) is involved in the process. 

The following general conclusions can be drawn about the properties of the angle-integrated probabilities and PADs: 
\begin{enumerate}

\item For the (a)- and (c)-geometry, the IR phase $\phi$ affects both the overall probability for an electron to be emitted at a given energy ($\sigma=\sigma(\phi)$) and the angular anisotropy parameters ($\beta=\beta(\phi)$);  all of the angular anisotropy parameters are real; PAD inheres the symmetries of the resulting field: for case (a) three orthogonal symmetry planes (see fig.~\ref{fig:3}a) and for case (c) axial symmetry with respect to the IR polarization vector accompanied with orthogonal symmetry plane (see fig.~\ref{fig:3}c).

\item For the (b)- and (d)-geometries, the phase-averaged part of eqs. (\ref{eq:b}) and (\ref{eq:d}) which contains $B^{dd}$ and $B^{uu}$ is axially symmetrical, while the interference term contains $B^{ud}$ and $B^{du}$ that depend on azimuth angle $\varphi$; as a result, a complete PAD possesses the only one symmetry plane orthogonal to the IR propagation direction. Neither angle-averaged spectrum nor the PAD changes with variation of the IR phase $\phi$ except for the rotation of the last around $z$-axis: eqs.~(\ref{eq:b2}) and (\ref{eq:d2}) incorporating $\beta_{22,42}$ depend on IR phase as $\exp[\pm 2i (\phi-\varphi)]$. The angular anisotropy parameters $\beta_{2(4),2}$ caused by interference  are complex.

\item The circular magnetic dichroism can be observed only for the (d)-geometry.

\item The angular anisotropy parameters being a ratio of harmonic functions of $\phi$ are periodical but not harmonic functions of IR phase $\phi$. 

\end{enumerate}

\subsection{Helium-like system}

\begin{figure}[h!]
\centering
\includegraphics[width=0.45\textwidth]{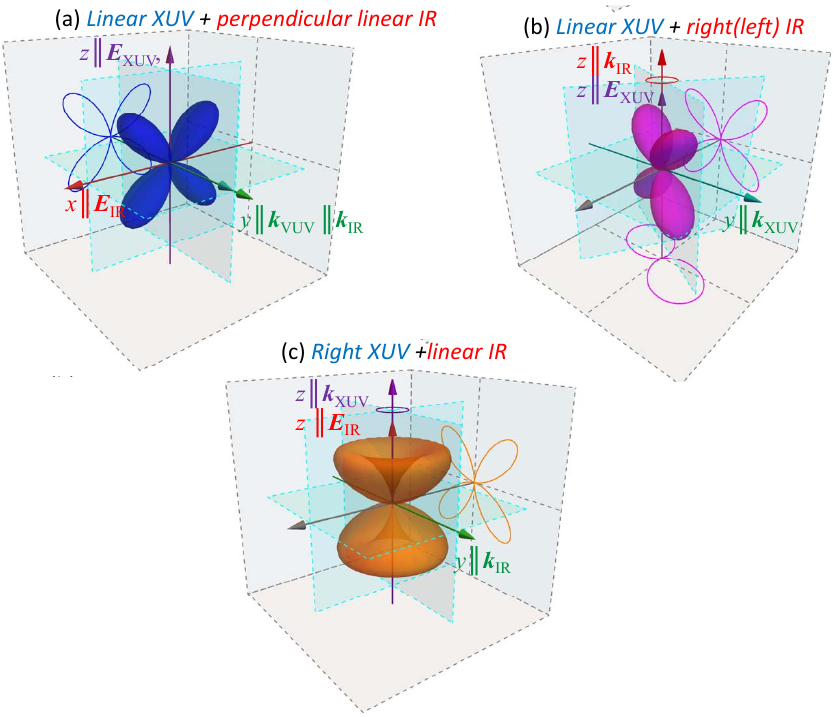} 
\caption{The sample of PAD for the case of helium-like system and different geometries.}
\label{fig:3}
\end{figure}

In this paragraph we consider additional features which arise when the ionized shell is a $s$-shell, so called {\it helium-like system}. In this case, the number of allowed ionization channels reduces significantly and they are characterized only  with photo\-electron angular momentum $l$. For (a)--(c) geometries, there is only one allowed channel: ionization to $d$-wave, for (d)-geometry, there are two channels: $d$- and $s$-waves (see table~\ref{tab1}).  Equation~(\ref{eq:W}) for the side\-bands turns  into an extremely simple form:

\begin{widetext}
\begin{eqnarray}
%W(\theta,\varphi)&=&\frac{1}{3}\sum_{kll'}(-1)^{(-m')}\CGC{l}{0}{l'}{0}{k}{0}\CGC{l}{m}{l'}{-m'}{k}{q}\,\nonumber\\
%&& \CGC{1}{\Lambda}{1}{\lambda}{l}{m}\CGC{1}{\Lambda}{1}{\lambda'}{l'}{m'}U^{(2)}_{\varepsilon l}U_{\varepsilon l'}^{(2)\ast}\sqrt{\frac{4\pi}{2k+1}}Y_{kq}(\theta,\varphi) \,.  \\
W(\theta,\varphi)^{(a)}&=&%\frac{-1}{6}\sum_{k\lambda\lambda'}\CGC{2}{0}{2}{0}{k}{0}\CGC{2}{\lambda}{2}{\lambda'}{k}{q}\frac{\sqrt{4\pi}}{\hat{k}}Y_{kq}(\theta,\varphi)  D_{\varepsilon d}^{u}D_{\varepsilon d}^{d,\ast}\,. \\
 \frac{1}{8\pi} (|D_{\varepsilon d}^{u}|^2+|D_{\varepsilon d}^{d}|^2+D_{\varepsilon d}^{u}D_{\varepsilon d}^{d\ast}+D_{\varepsilon d}^{d}D_{\varepsilon d}^{u\ast}) \sin^2\theta\cos^2\theta\cos^2(\varphi-\phi-\gamma) \label{eq:S1}\\
 %\widetilde{W}(\theta,\varphi)^{(a)}&=&\frac{-1}{6}\sum_{k\lambda\lambda'}\CGC{2}{0}{2}{0}{k}{0}\CGC{2}{\lambda}{2}{\lambda'}{k}{q}  2\Re [U_{\varepsilon d}^{u}D_{\varepsilon d}^{d,\ast}]\frac{\sqrt{4\pi}}{\hat{k}}Y_{kq}(\theta,\varphi)\,. \\
% &=&\frac{1}{4} \sin^2\theta\cos^2\theta  2 \sin(\varphi-\phi)\Re[e^{i(\varphi-\phi+\pi/2)}U_{\varepsilon d}^{u}D_{\varepsilon d}^{d,\ast}]\,.\\
 W(\theta,\varphi)^{(b)}&=&%\frac{-1}{6}\sum_{k}\CGC{2}{0}{2}{0}{k}{0}\CGC{2}{1}{2}{-1}{k}{0}  (|D_{\varepsilon d}^{u}|^2+|D_{\varepsilon d}^{d}|^2) P_{k}(\cos\theta)\,. \\
 \frac{1}{16\pi} (|D_{\varepsilon d}^{u}|^2+|D_{\varepsilon d}^{d}|^2+e^{2i\varphi} D_{\varepsilon d}^{u}D_{\varepsilon d}^{d\ast}+e^{-2i\varphi} D_{\varepsilon d}^{d}D_{\varepsilon d}^{u\ast})\cos^2\theta \sin^2\theta \,.\label{eq:S2} \\
 W(\theta,\varphi)^{(c)}&=&%\frac{-1}{6}\sum_{k}\CGC{2}{0}{2}{0}{k}{0}\CGC{2}{1}{2}{1}{k}{2}\frac{\sqrt{4\pi}}{\hat{k}}Y_{kq}(\theta,\varphi)  D_{\varepsilon d}^{u} D_{\varepsilon d}^{d,\ast}\,. \\
% \frac{1}{4} \sin^2\theta\cos^2\theta  2\Re[e^{2i\varphi} D_{\varepsilon d}^{u}D_{\varepsilon d}^{d,\ast}]\,. \\
% W(\theta,\varphi)^{(c)}&=&%\frac{-1}{6}\sum_{k}\CGC{2}{0}{2}{0}{k}{0}\CGC{2}{1}{2}{1}{k}{0}P_{k}(\cos\theta)  d_{\varepsilon d}^{u}d_{\varepsilon d}^{d,\ast}\,. \\
 \frac{1}{16\pi}(|D_{\varepsilon d}^{u}|^2+|D_{\varepsilon d}^{d}|^2+D_{\varepsilon d}^{u}D_{\varepsilon d}^{d\ast}+D_{\varepsilon d}^{d}D_{\varepsilon d}^{u\ast}) \cos^2\theta\sin^2\theta \,;\label{eq:S3} \\
W(\theta,\varphi)^{(d)}&=&\frac{1}{12\pi}\sum_{kll'} \CGC{l}{0}{l'}{0}{k}{0}^2  \CGC{1}{1}{1}{-1}{l}{0}\CGC{1}{1}{1}{-1}{l'}{0} D^d_{\varepsilon l} D_{\varepsilon l'}^{d\ast} P_{k}(\cos\theta)+\frac{1}{32\pi} |D^u_{\varepsilon d}|^2 \sin^4\theta \,\nonumber\\
&-&\frac{1}{12\pi}\sum_{kl'}\CGC{2}{0}{l'}{0}{k}{0}\CGC{2}{2}{l'}{0}{k}{2} \CGC{1}{1}{1}{-1}{l'}{0}\frac{\sqrt{4\pi}}{\hat{k}} (D_{\varepsilon d}^uD_{\varepsilon l'}^{d\ast}Y_{k2}(\theta,\varphi)+D_{\varepsilon d}^{d\ast}D_{\varepsilon l'}^{d}Y_{k-2}(\theta,\varphi))\,. \label{eq:s4} 
\end{eqnarray}
\end{widetext}
Here $D^{u/d}_{\varepsilon l}\equiv D^{(2),u/d}_{(0l)l}$. In figure \ref{fig:3}, the general pattern of PAD for helium-like system is presented. For the (a) and (c)-geometries they are unconditional, for (b) plotted under assumption that $D^{u}_{\varepsilon d}=D^{d}_{\varepsilon d}$.

In the case of helium-like system one may conclude the following: 
\begin{enumerate}
\item For the (a) and (c)-geometries, only $d$-wave left, and absorption and emission amplitudes come into the equations equally. The PADs turn into completely geometrical form with  $\beta^{(a),(c)}_2=5/7$, $\beta^{(a),(c)}_4=-12/7$, $
\beta^{(a)}_{22}=-5\sqrt{6}/7$, and $\beta^{(a)}_{42}=  -6\sqrt{10}/7$. 

%\begin{eqnarray}
% \beta^{(a),(c)}_2&\hspace{-20pt}=\dfrac{5}{7}&\hspace{10pt}\beta^{(a),(c)}_4=  -\frac{12}{7}  \\
%\beta^{(a)}_{22}&=-\dfrac{5\sqrt{6}}{7}&\hspace{10pt}\beta^{(a)}_{42}=  -\frac{6\sqrt{10}}{7} 
%\end{eqnarray}
%%\begin{align*}
 %%\beta^{(a),(c)}_2&=\dfrac{5}{7}&\beta^{(a),(c)}_4&=-\frac{12}{7}  \\
%%\beta^{(a)}_{22}&=-\dfrac{5\sqrt{6}}{7}&\hspace{10pt}\beta^{(a)}_{42}&=  -\frac{6\sqrt{10}}{7} 
%%\end{align*}
\item For the (b)-geometry, $d$-wave absorption ($^u$) and emission ($^d$) amplitudes contribute to PAD  differently, and the PAD is partly geometrical: $\beta^{(b)}_2=5/7$, $\beta^{(b)}_4=  -12/7$,
\begin{align*}
 %\beta^{(b)}_2&=\dfrac{5}{7}&\beta^{(b)}_4&=  -\frac{12}{7}  \\
 \beta^{(b)}_{22}&=\frac{5\sqrt{6}D^u_{\varepsilon d}D^{d\ast}_{\varepsilon d}}{7(|D^u_{\varepsilon d}|^2+|D^d_{\varepsilon d}|^2)}&\beta^{(b)}_{42}&=\frac{6\sqrt{10}D^u_{\varepsilon d}D^{d\ast}_{\varepsilon d}}{7(|D^u_{\varepsilon d}|^2+|D^d_{\varepsilon d}|^2)}.
\end{align*}

\item For (a)--(c) geometries, the maximal probability of electron emission is observed at the polar angle $\theta=\pi/4$.

\item For (c)- and (d)-geometries, PAD possesses the same symmetries as in the general case. For (a)- and (b)-geometries, two additional symmetry planes arise: for (a) they are defined by the geometry and make angle $\pm\pi/4$ with polarization vectors of IR and XUV comb; for (b) they defined by phase between up- and down-pathways (${\rm Arg}[D^u_{\varepsilon d}D^{d\ast}_{\varepsilon d}]$).

\item For the (d)-geometry, there is a difference between pathways with the absorption and emission of an IR photon. While the first one leads to $d$-wave only, the second allows also $s$-wave. Therefore, the absorption pathway contributes to the plane orthogonal to the fields' propagation direction, and the emission pathway may contribute in the fields' propagation direction.
\end{enumerate}

%good place 

A short comment should be given about {\it circular dichroism} in the scheme (d). In this geometry, there is an essential difference in the allowed channels for the case when IR and XUV components have the same helicity compared to when they have opposite helicity. In the first case, pathways with IR absorption lead to $L=2$ ($\varepsilon d$ for helium) , while those with emission lead to $L=0\,,1\,,2$ ($\varepsilon s(\varepsilon d)$ for helium). In the second case, the situation is opposite. One can easily cast the equation for dichroism using (\ref{eq:d}) and (\ref{eq:s4}). The interesting feature is that contribution of fourth-rank terms are strictly canceled in the dichroism for any target.  It could be important for experimenters because extraction of higher rank anisotropy parameters are usually more difficult than those of 2nd rank.

Nevertheless for the region of smooth continuum chosen for numerical calculations in the next section, the difference in probabilities of up- and down-channels is little, and dichroism is not significant. One should look for a system with a sharp spectroscopic feature (auto\-ionizing resonance) to observe the circular magnetic dichroism in a such setup.

%\begin{widetext}
%\begin{eqnarray}
%\mathcal D(\theta,\varphi)^{(d)}&=&\frac{1}{4\pi}(\frac{`}{9}(|D^d_{\varepsilon s}|^2+|D^d_{\varepsilon d}|^2-|D^u_{\varepsilon s}|^2-|D^u_{\varepsilon d}|^2)+\frac{1}{6} (|D^u_{\varepsilon d}|^2-|D^d_{\varepsilon d}|^2) \cos^2\theta+
%\frac{1}{9\sqrt{2}}D^d_{\varepsilon l} D_{\varepsilon l'}^{d\ast} P_{k}(\cos\theta) \,\nonumber\\
%&-&\frac{1}{6\sqrt{2}} \sin^2\theta(D_{\varepsilon d}^uD_{\varepsilon l'}^{d\ast}\exp[2i\varphi]+D_{\varepsilon d}^{u\ast}D_{\varepsilon l'}^{d}\exp[-2i\varphi])
%\end{eqnarray}
%\end{widetext}
%%%%%%%%%%%%%%%%%%%%%%%%%%%%%%%%%%%%%%%%%%
\section{\label{sec:2} Numerical calculations and Discussion}

\begin{figure}
\centering
\includegraphics[width=0.45\textwidth]{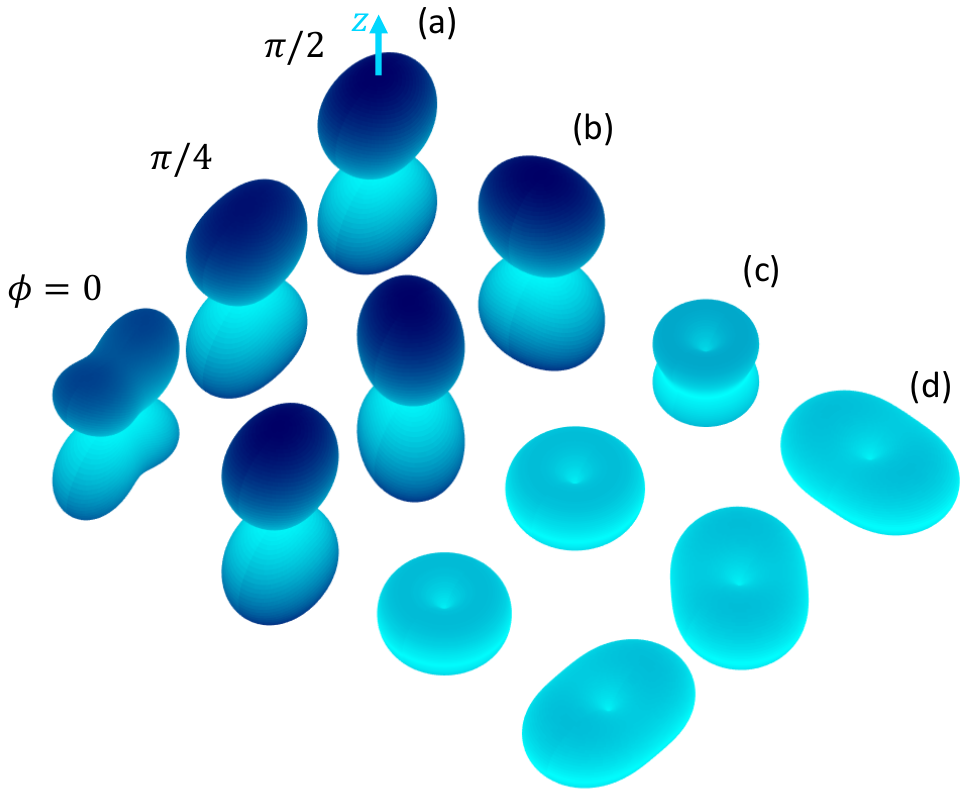} 
\caption{PADs as defined in eqs. (\ref{eq:a})--(\ref{eq:d2}) for geometries (a)--(d), respectively, but normalized to $\sigma=1$ for IR field phase $\phi=0,\pi/4,\pi/2$ for the side\-band at $\approx9.5$ eV.}
\label{fig:results_neon_pad}
\end{figure}

In this part, the results of calculations of photo\-ionization probability and angular anisotropy parameters in neon valance shell ionization by the pulse consisting of IR with $\omega=1.55$~eV and its 17th, 19th and 21nd harmonics are presented. The electric field parameters were set to the values typical for RABBITT experiments: $E_{\rm{xuv}}^0=10^{-4}$ a.u., $E_{\rm{ir}}^0=2.5\cdot10^{-3}$ a.u., and $\tau=10$~fs. Note that according to eq.~(\ref{eq:field}), duration of XUV components is twice shorter than one of IR component.

Reduced dipole matrix elements between  ground state $2s^22p^{6\, 1}S$ and continuum states $2s^22p^5\varepsilon l ^{\,1\!}L$ were calculated in MCHF package \cite{Fischer1997} with non-orthogonal $2p$ orbital. For the ground state, the experimental ionization energy was used. Reduced dipole matrix elements between continuum states were calculated using a method described in \cite{Mercouris1996} involving  angular momentum algebra \cite{popova2023} to convert the radial integrals to the matrix elements in an appropriate angular momentum coupling scheme.

In figure \ref{fig:results_neon}, there are dimensionless parameters $\beta_{kq}$ defined by eqs. (\ref{eq:a})--(\ref{eq:d2}) calculated by both RE and PT methods plotted together with integrated spectrum. Here we set all $\phi_{N}$ to zero. The central peak is main\-line caused by 19th harmonic, two lower lines on the sides of ML19 are side\-bands SB18 and SB20. For (a)- and (c)-geometry,  calculations at three phases $\phi=0\,,\pi/4$ and $\pi/2$ of the IR component are presented. For (b) and (d)-geometry, presented parameters do not depend on IR phase. For the (a)-geometry at $\phi=0$ and for (c)-geometry at $\phi=\pi/2$, the integrated probability is negligible, therefore dimensionless anisotropy parameters are much less trustworthy and are not presented.  
The overall agreement between the methods is good. As RE calculation is more rugged, it is  shown partly where it is necessary for discussion and do not interfere the total readability.

% The most significant differences observed for a phase $\phi$ corresponding to minimal photo\-emission probability at the side\-bands, for example, between dotted lines and crosseы in panel (c). It is natural since $\beta$ parameters are defined as fraction with cross section in the denominator. 

%Because of the channels involved differently to the zero,   second and fourth ranks  IR-delay affects the form too, not simple scaling or rotation. Because the same reason for (b)- and (d)-geometries  the symmetry planes defined by the phase are not relevant.

The angular anisotropy parameters vary smoothly through the lines except the edges where photo\-emission probability drops down. As it is clear from equations (\ref{eq:b2}) and (\ref{eq:d2}), $\beta_{22(42)}$ are responsible for the PADs' dependency on azimuth angle, while $\beta_{2(4)}$ are responsible for axially symmetrical contribution. In all the cases, parameters $\beta_{22}$ are much smaller than $\beta_{2}$ and compatible with $\beta_{4}$ and $\beta_{42}$. The minor value of these parameters is a result of interference between the ionization to $f$- and $p$-wave. 

For the geometries (a)--(c), if  only $f$-wave is important due to a spectroscopic feature, the anisotropy parameters tend to $\beta_2\rightarrow 4/7$, $\beta_4\rightarrow -4/7$, $\beta_{22}\rightarrow -2\sqrt{6}/7$ (if exists) and $\beta_{42}\rightarrow -\sqrt{10}/7$ (if exists). 

As it was already mentioned, in the (a) and (c) geometries, the spectrum as well as parameters $\beta_{kq}$ depend on the IR phase $\phi$ (see panels (a) and (c) in figure \ref{fig:results_neon}). For the (a)- and (c)-geometry, formal equations for the integrated spectrum and $\beta_4$ in terms of dynamical parameters (\ref{eq:Buu}) coincide.  Nevertheless, one should remember that the corresponding amplitudes depend on IR phase, and the last being written in form of eq. (\ref{eq:field}) possesses a different physical meaning for different geometries. The easiest way to illustrate is that $\phi=0$ means $\bm{E}_{\rm{xuv}}\parallel z$ and $\bm{E}_{\rm{IR}}\parallel x$ for (a)-geometry, while  $\phi=0$ means $\bm{E}_{\rm{xuv}}\parallel x$ and $\bm{E}_{\rm{IR}}\parallel y$ for (c)-geometry. Therefore, the same values of observables are reached at different phases. The difference in $\beta_2$ originates from different signs before the interference terms $B^{\nu\mu}[1,2]$ (see factor $(-1)^{L+L'+1}$ in eq. (\ref{eq:a})). If for some reason either $L=1$ ($P$-term) or $L=2$ ($D$-term) dominates, $\beta_2$ in these geometries would also coincide. 

\begin{figure}[t!]
\begin{minipage}[c]{\textwidth}
\centering
\includegraphics[width=1.0\textwidth]{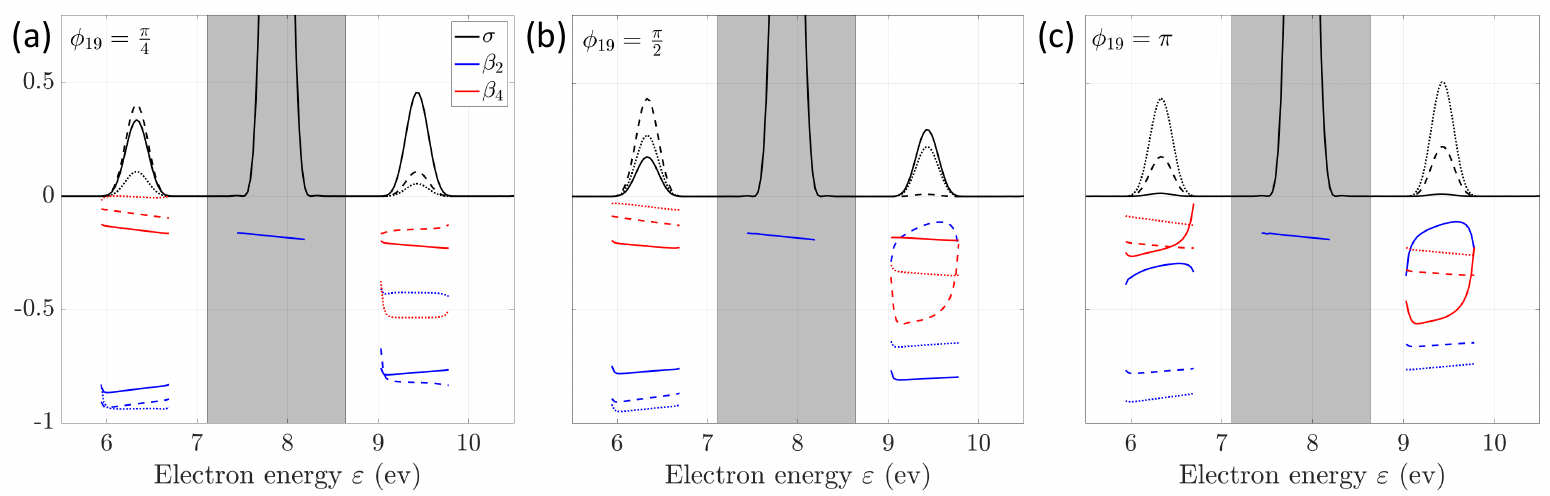} 
\caption{PT calculations as in figure \ref{fig:results_neon}c, but for different phases of 19th harmonic.}
\label{fig:phase}
\end{minipage}

\end{figure}

In figure \ref{fig:results_neon_pad}, the calculated PADs for different geometries and IR phases are presented.  In the schemes (a) and (b), $\beta_2$ is positive, therefore, the maximum photo\-electron emission is found along the quantization axis ($\theta=0,\pi$) because the other $\beta$s are much smaller.  $\beta_2$ is also positive when both of the field components are linearly polarized in the same direction \cite{Joseph2020, yudin2023}. In the schemes (c) and (d), $\beta_2$ is negative, therefore, the maximum photo\-electron emission is found in the plane perpendicalar to the quantization axis ($\theta=\pi/2$). The PADs keep their form for the phases corresponding essential signals in the side\-bands and diverge for the phases corresponding a minor side\-bands  ($\phi=0$ for (a)- and $\phi=\pi/2$ for (c)-geometry) they become quite diverse: as $\beta_{2}$ sharply changes (its absolute values decrease, similar to fig.~\ref{fig:phase}), $Y_{4,22,42}(\theta,\varphi)$ play a more significant role (see fig. \ref{fig:results_neon_pad}a,c). For these phases, PADs resemble the ones for helium-like case (fig. \ref{fig:3}a,c).
In schemes (b) and (d), as it was previously mentioned, PAD rotates around $z$-axis,  with variation of the IR component phase $\phi$. For neon the effect of rotation is more prominent for the geometry (d) (see fig. \ref{fig:results_neon_pad}b,d).

Until now, we assumed that all of the XUV harmonics phases are zero $\phi_{N}=0$. In order to investigate the role of the XUV phases, we chose (c)-geometry (see fig.~\ref{fig:phase}) and assigned several different values to $\phi_{19}$. One can see that varying the phase of an XUV component leads to the re\-arrangement of the photo\-emission signal therefore the same magnitudes are achieved at different IR phases.  The range of variation for the anisotropy parameters remains the same as well.

\section{Conclusion}

In the paper we investigated how the polarization and propagation direction of the field's components affect kinematic of photoionization in the RABBITT scheme. We considered: (a) crossed linearly polarized IR and XUV harmonics; (b,c) either XUV or IR is linearly polarized and the remaining component is circularly polarized (d) both IR and XUV are circularly polarized.

Among the considered geometries, the setup with circularly polarized XUV harmonics and linearly polarized IR component possesses the highest symmetry (axial). This geometry as well as setup with crossed linearly polarized components, allows for the observation of RABBITT oscillations in both angle-integrated  and angle-resolved probabilities of electron emission. 

On the contrary, for a circularly polarized IR field and either linearly or circularly polarized XUV harmonics, there is only one symmetry plane, and RABBITT oscillations are observed only in the angular distribution of photo\-emission. In these geometries,  the variation of the IR phase appears as a  rotation of PAD with respect to axis oriented along the direction of IR component propagation.

To distinguish between geometrical (inherited solely from the polarization of the electromagnetic field) and spectroscopic (inherited from the properties of the target, and thus dependent on photon energies) features, we considered neon and helium-like targets. In the first case, the observable values are determined by the interplay between different ionization channels, whereas for helium, the system may reduce to completely geometrical,  with a PAD that does not depend on dynamic parameters, such as photon energy or even specific atom. 

For circularly polarized XUV comb, the symmetries are the same for both multichannel targets and helium-like ones. In geometries with linearly polarized XUV component, there are two additional symmetry planes for helium-like targets, moreover, for linearly polarized IR component, they are geometrical and make angle $\pm\pi/4$ with polarization vector, while for circularly polarized IR component, they are dynamical and depend on phase between up- and down-transitions.

\section*{ACKNOWLEDGMENTS}
The part of research related to the general approach (section II) was funded by the Russian Science Foundation (RSF) under project No. 22–12–00389 and the numerical simulations were supported  Ministry of Science and Higher Education of the Russian Federation grant No. 075-15-2021-1353.


%=====================================
% References, variant A: internal bibliography
%=====================================

%\nocite{*}
\bibliography{references}

\end{document}